\documentclass[aps,pra,twocolumn,showpacs,floatfix]{revtex4-1}

\usepackage{dsfont}
\usepackage[dvips]{graphicx}
\usepackage[english]{babel}
\usepackage{amsmath,hyperref}
\usepackage{amssymb}
\usepackage{slashed}

\newif\ifhyper
\hypertrue
\ifhyper
\hypersetup{
  citecolor = {green},
  colorlinks = {true}, 
  urlcolor = {blue} 
}

\newcommand{\mean}[1]{\langle #1 \rangle}

\def\eps{\epsilon}

\def\w{\omega}

\def\calD{{\cal D}}

\begin{document}
\title{Equilibrating dynamics in quenched Bose gases: characterizing
multiple time regimes }

\author{A. Ran\c con and K. Levin}

\affiliation{James Franck Institute and Department of Physics,
University of Chicago, Chicago, Illinois 60637, USA
}

\begin{abstract} 
We address the
physics of equilibration in ultracold atomic gases following
a quench of the interaction parameter.
Our work is based on a bath model which generates
damping of the bosonic excitations. We illustrate this
dissipative behavior through
the
momentum distribution of the excitations, 
$n_k$,
observing that
larger $k$ modes have shorter relaxation times $\tau(k)$; they
will equilibrate faster, as has been claimed in
recent experimental work.
We identify three time regimes. At short times
$n_k$ exhibits oscillations; these are damped out at intermediate
times where the system appears to be
in a false or slowly converging equilibrium. Finally, at longer times, full equilibration occurs. 
This false-equilibrium
is, importantly, associated with the $k$ dependence in $\tau(k)$
and has implications for experiment.
\end{abstract}
\pacs{03.75.Kk, 98.80.-k, 47.37.+q, 43.20.Ks}
\maketitle

\textit{Introduction-}
Recent interaction quench experiments in cold bosonic gases are
providing unique perspectives into the behavior of out-of-equilibrium dynamics of quantum
systems \cite{Donley2001,Greiner2005,Smith2012,Hung2013,Makotyn2014}. 
These perspectives were hitherto not available in the quantum fluids of
condensed matter. The extent to which equilibrium is accessible and the time constants for
equilibration are all open questions.
Equally of interest is the nature of metastable states, (often) so produced.
Considerable theoretical attention 
has gone into this subject, albeit 
characterizing the post-quench physics
entirely in terms of oscillatory behavior \cite{Barankov2004,Altschuler2,Natu2013,Yin2013,Sykes2014,Kain2014}.

How long do these oscillations persist and
how does ultimate equilibration proceed for different momentum
states is a
complicated problem that is the focus of the present paper.
Here we discuss the different time scales
associated with
dissipation and equilibration in the context of the evolution of the momentum distribution
$n_k$ for a three-dimensional Bose gas. 
While we use a specific bath model to derive detailed results for $n_{k}(t)$, our central results
can almost be anticipated by making use of empirical observations in previous
quench experiments \cite{Hung2013,Makotyn2014}.
As emphasized in both experiments, the equilibration dynamics is rather strongly dependent
on the momentum of the state under consideration.
An unpublished analysis 
\cite*{[{See discussion in Ref. 24 in the preprint version of Ref. 4: }] arxiv} of the experiments in Ref.~\onlinecite{Hung2013},
led to the conclusion that damping at
large momentum had to be included. Also notable is the claim that
``it is perhaps not unexpected that higher momenta dynamics
saturate faster" \cite{Makotyn2014}.
This demonstration that 
large momentum $k$, high energy, states
equilibrate more rapidly than those at small $k$ is the aim of this
paper. It leads to
a multi-step equilibration process, assuming,
as is reasonable, that the condensate also evolves
in time as the system re-equilibrates.

The important point at issue is that
the relaxation times $\tau(k)$ 
disperse with $k$. At
some intermediate time after the quench, there will always be higher energy
$k$ states which will be able to follow quasi-adiabatically the 
(necessarily) slower
relaxation of the condensate. 
But lower energy states, as well as the condensate,
will not yet have equilibrated.
This suggests, as has been claimed in the literature \cite{Smith2012} that,
after the initial time period in which $n_k$ oscillates,
there will be a two stage 
equilibration process, associated with the false equilibrium
of the
high $k$ states, and the ultimate true equilibration of the full system.

The fact that large $k$ is observed \cite{Makotyn2014}
to equilibrate first suggests that theoretical calculations of the short
distance behavior should not be characterized entirely
by an oscillatory time dependence as might be associated with
short time evolution. Arriving at an understanding of
the short distance behavior 
should also include dissipation mechanisms.
More generally, a description of the post quench behavior
entirely in terms of non-dissipative oscillatory contributions (although they may not
be that apparent after integration over momentum) 
is argued here to be inadequate.
In this regard, we differ from the literature \cite{Natu2013,Yin2013,Sykes2014}.

In this paper we focus on including this dissipation and will
demonstrate that the $k$ dependence claimed in Ref.~\onlinecite{Makotyn2014}
is consistent with our calculations.
We will use a simple bath model \cite{Rancon2013b}, but
before doing so, we begin at a more heuristic level,
using another point of view, that of 
 the dissipative versions 
of the 
Gross-Pitaevski equation (DGPE) \cite{Choi1998,Stoof1999,Zaremba1999,Blakie2008}. 
It should be stressed that the subject matter of this paper does not
concern the DGPE as such, except as a back-of-the-envelope
method for arriving at the results of the bath model.
Nevertheless, this approach will give us the prototypical time-scales, showing the generality of our arguments.
In these approaches, the equation of motion of the (mostly condensate)
field is given by
$i\partial_t \phi(x,t)=$ 
\begin{equation}
\big[1-i\gamma(x)\big]\bigg\{-\frac{\nabla^2}{2m}-\mu+V(x)+g|\phi(x,t)|^2\bigg\}\phi(x,t),
\label{eq_SPGE}
\end{equation}
where $V(x)$ is the trap potential and $g$ the two-body interaction strength. 
Here $\gamma$ describes the dissipation processes 
and its specific form depends on the model used to derive the DGPE. 
 Here and in the following, we work in units such that $\hbar=k_B=1$.

One can deduce some simple physical results from this dissipative
GP equation. Throughout we ignore trap effects as has been
argued  to be appropriate for
times smaller than the inverse trap frequency \cite{Hung2013}. The DGPE  for a perturbation $\delta \phi_k$  from the equilibrium solution 
$\phi_0$
is schematically of the form
\begin{equation}
i  \partial_t \delta\phi_k = (1-i \gamma)\Big[ (\epsilon_k+g n_0) \delta\phi_k+ g n_0 \delta\phi_k^*\Big],
\end{equation}
where $\eps_k=\frac{k^2}{2m}$, the condensate $n_0=|\phi_0|^2$, and  
$\delta\phi_k$
 represents an excited state having
momentum $k$. 
Thus it will qualitatively behave as
\begin{equation}
\delta\phi_k(t)\propto e^{-i \sqrt{E_k^2-(\gamma gn_0)^2 } t -\gamma (\epsilon_k+g n_0) t},
\label{eq:3}
\end{equation}
with Bogoliubov energy $E_k=\sqrt{\eps_k\big(\eps_k+2 g
n_0\big)}$.
This simple analysis shows that
there are two distinct time dependences: an oscillatory contribution 
which is proportional to the energy $E_k$ (for sufficiently large momentum) and
a
damping contribution which scales with the energy $\w_k=\epsilon_k+gn_0$, and
is multiplied by a dissipative factor $\gamma$ as well.

To make these heuristic arguments we presume (for the moment)
that the condensate $n_0$ has little or no time dependence.
Under this assumption, we may read off from 
Eq.~(\ref{eq:3})
the relaxation time associated with
the damping of oscillations
$$\tau_{\rm interm} (k) \propto \frac{1}{\gamma \w_{k}}.$$
Importantly, from this equation we note that higher energy
or larger $k$ modes will equilibrate faster.
By contrast, we associate the short time (\emph{i.e.} undamped) dynamics 
with the characteristic time
$$\tau_{\rm short}(k) \propto\frac{1}{E_k}.$$
provided $k$ is sufficiently large \footnote{In all rigor, we find $\tau_{\rm short}(k) \propto\frac{1}{\sqrt{E_k^2-(\gamma gn_0)^2 }}$. For momentum smaller than a cross-over momentum $ k_x$ (defined such that $E_{k_x} \equiv \gamma gn_0$),  the dynamics of the excitations is very different 
from that of (undamped) Bogoliubov theory.}.
This typical time-scale is
associated with the oscillation period of observables as predicted by Bogoliubov theory \cite{Natu2013,Rancon2013b,Yin2013,Kain2014,Sykes2014}.

We see that as long as 
$t \ll  \tau_{\rm interm} (k)$, 
we can ignore the damping, and the system behaves as if it were described by Bogoliubov theory and its variants. 
Here there is always a range in time where the short time undamped
dynamics is correct, but this range gets 
smaller and smaller as $k$ increases. 
However, this undamped dynamics is meaningful only if there is a clear separation of scales $\frac{\tau_{\rm short}(k)}{\tau_{\rm interm }(k)}
\simeq\gamma\frac{\w_k}{E_k}\ll1$.
This also implies that if one works at fixed time  
as is done in the experiments (call that time $t_{\rm exp}$), the short time dynamics 
will not be able to describe the physics for momenta such 
that $\tau_{\rm interm} (k)<t_{\rm exp}$.
This has implications for extracting the Tan contact parameter \cite{Tan2008}.
 
\begin{figure}
\includegraphics[width=2.5in,clip]
{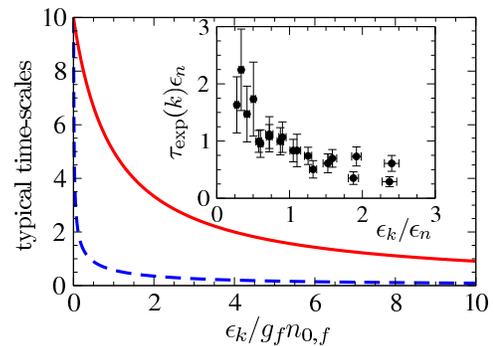}
\caption{Characteristic times $\tau_{\rm short}(k)$ (dashed line) and  $\tau_{\rm interm}(k)$ (solid line) in units of $(g_f n_{0,f})^{-1}$ vs $\epsilon_k/g_f n_{0,f}$ ($\gamma=0.1$, see text). Inset: data from Ref. \onlinecite{Makotyn2014} showing the time-scale $\tau_{\rm exp}(k)$ after which the experimental system has attained a steady state for each momentum $k$. $\epsilon_n=(6\pi^2n)^{2/3}/2m$ correspond to the `Fermi energy' (with $n$ the density of atoms), which is the only  energy scale for a unitary Bose gas. 
The 
specific parameters for these plots are 
discussed in the context of the subsequent figures. The abscissas units of the graph and the inset are compared in the text.}
\label{fig_tk}
\end{figure} 
 
Figure \ref{fig_tk} shows the typical behavior for the short
$\tau_{\rm short}(k)$
(dashed line) and
intermediate
$\tau_{\rm interm}(k)$
(solid) relaxation times.
Both
are peaked at small $k$, and
equilibration is very fast at
short distances. The inset is from Ref. \onlinecite{Makotyn2014}.
This inset sets up the underlying experimental challenge
that 
``the momentum dependence of the
time scales
remains to be understood". This will be discussed in more detail below.

In the above heuristic argument we have not considered the possibility that
the relaxational dynamics of the condensate may contribute
an additional time-scale.
Indeed, because of the wide spread, associated with
the $k$-dependence of the excitation relaxation
times, 
the full equilibration process is more complex. 

At a simple level we can characterize the 
time dependence of the condensate in terms of a single relaxation time
$\gamma_0^{-1}$
\begin{equation} n_0(t)=n_{0,f}+ h(\gamma_0
t)(n_{0,i}-n_{0,f}), \label{eq_n0}
\end{equation} where we introduce a damping function
$h(\gamma_0 t)$ which, for example, can be taken as a simple
exponential, $e^{-\gamma_0 t}$. Here $n_{0,i}$
($n_{0,f}$) is the initial (final) value of the condensate,
associated with a quench.

The details of this phenomenology are in no way essential to the
arguments in this paper. We present it here for illustration
purposes.
Indeed, one could contemplate very non-monotonic functional forms.
What is essential here is rather the behavior at the end-stage
of condensate evolution, where presumably the condensate approaches
equilibrium in a monotonic fashion.
This condensate evolution represents another time-scale in the equilibration
process
$$\tau_{\rm long } \propto \gamma_0^{-1}. $$
For sufficiently large momentum, we have $\tau_{\rm interm}(k)\ll \tau_{\rm long }$, implying that the high energy
modes will equilibrate faster than the condensate.

In the spirit of simplicity, we adopt Eq (\ref{eq_n0}),
with an exponential damping function, for definiteness.
Here we presume that
after an interaction quench,
particularly near unitarity as in Ref.  \onlinecite{Makotyn2014},
the condensate density $n_0(t)$ evolves and most probably
decreases as the system reaches a new equilibrium state.

We stress the contrast here with
fermionic superfluids, where on the basis of Bogoliubov-de Gennes theory,
the order parameter dynamics is obtained
from the excitation dynamics. Indeed, it has been argued
that the
same should apply to quenched Bose gases, through use of
the number equation \cite{Yin2013,Kain2014}. In this way the
condensate dynamics 
are derived from, and therefore somewhat
secondary to that of the excitations.
Our emphasis in this paper is on the inclusion of dissipation
which is presumably rather independent of the condensate
(and more directly associated with higher energy states not
included in Bogoliubov theory).
The spirit here is closer to that of conventional bosonic
Bogoliubov theory
where
the condensate has an intrinsic dynamics, distinct from that of the
excited states.

\textit{Overview of Bath Approach-}
We now 
characterize the time evolution of the equilibration
dynamics concretely through the study of the momentum distribution $n_k(t)$ using a bath model. We
implement 
quantum dissipation following the work of
Caldeira and Leggett \cite{Caldeira1983}. 
In this seminal paper \cite{Caldeira1983},
dissipation was induced by coupling the system to a bath
composed of an infinite set of harmonic oscillators. 
A connection between
the DGPE approach and that of Ref. \onlinecite{Caldeira1983}
was proposed by Stoof \cite{Stoof1999}.
If the system is
either a free particle or a particle confined in an harmonic oscillator, the
Hamiltonian is quadratic and one can solve the equations of motion exactly.

The introduction of the bath, as well as its parameters, 
has to be seen as mainly phenomenological. Nevertheless, the
bath is often viewed as reflecting
the incoherent (high energy) modes that are integrated out in other approaches (such as the higher-harmonics modes of 
the trap in the stochastic GPE approaches). These
allow energy to dissipate. The bath can be thought of as
incorporating the interactions between the different modes of the full many-body interacting system that would allow equilibration if treated beyond mean-field (Bogoliubov theory). 

The coupling between these extra degrees of freedom and the Bogoliubov modes is characterized by the so-called spectral function of the bath $\Sigma_2(\w)$. For an Ohmic bath \cite{WeissBook}, $\Sigma_2(\w)\propto \Gamma \w$, where $\Gamma$ describes the strength of the coupling that plays a similar role to that of $\gamma$ in Eq. \eqref{eq_SPGE} (for $\Gamma=0$, one recovers Bogoliubov theory and the results of Ref. \cite{Natu2013}). In our previous study \cite{Rancon2013b} of the quench dynamics of a two-dimensional Bose gas, we have shown that a value $\Gamma\simeq 0.1$ was consistent with the experiment of Ref. \onlinecite{Hung2013}. This is also the value that we will use here in our numerical results to illustrate the equilibration dynamics \footnote{More generally, these and the measurements
of $\tau(k)$ in Ref. \onlinecite{Makotyn2014} seem compatible with 
$\Gamma$ between $0.1$ and $1.0$.}.

The technical details of our calculation, as well as the explicit expression for the momentum distribution are given in the Supplemental Materials. The initial conditions corresponds to that of an ideal Bose gas at zero temperature that is quenched to a finite interaction strength.

\begin{figure}
\includegraphics[width=2.8in,clip]
{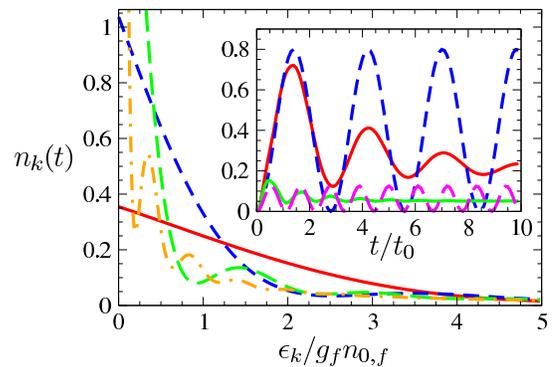}
\caption{$n_k(t)$ versus momentum, in terms of $\eps_k/gn_{0,f}$ , 
for four different times $t=t_0/2$ (solid curve), $t=t_0$ (short dashed curve), $t=2t_0$ (long dashed curve) and $t=5t_0$ (dot-dashed curve) for a 
time independent condensate $n_0(t)=n_{0,f}$. and $\Gamma=0.1$ 
The inset plots 
$n_k(t)$ versus $t/t_0$ for energy $\epsilon_{k}=g_f n_{0,f}$ (top solid curve) and $\epsilon_{k}=3g_f n_{0,f}$ (bottom solid curve), as well as  Bogoliubov results (\emph{i.e.} $\Gamma=0$) for the same energies (dashed curves).
}
\label{fig_energy}
\end{figure}

\begin{figure*}
\includegraphics[width=2.5in,clip]
{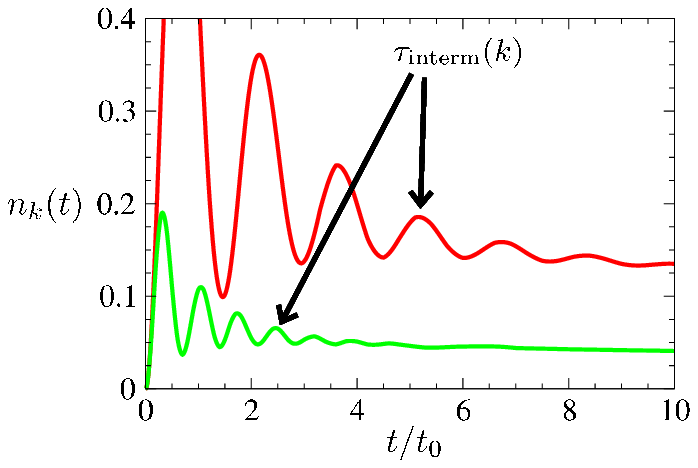}
\includegraphics[width=2.2in,clip]
{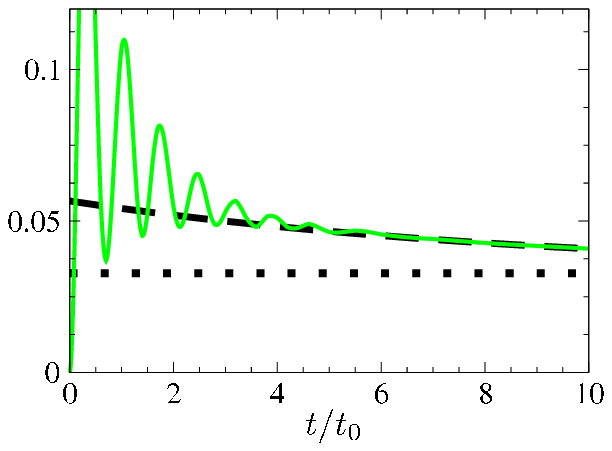}
\caption{Left:
$n_k(t)$ versus $t/t_0$ for a time dependent condensate (Eq. \eqref{eq_n0})
with $n_{0,i}=2n_{0,f}$ for $\epsilon_{k}=g_f n_{0,f}$ (top solid curve) and $\epsilon_{k}=3g_f n_{0,f}$ (bottom solid curve). The arrows show the typical damping time of the oscillations $\tau_{\rm interm}(k)$ for both energies.
 Right: Zoom on time-evolution of the $n_{k}(t)$ for $\epsilon_{k}=3g_f n_{0,f}$, showing in more detail the  long-time equilibration of $n_k(t)$  with
time-dependent condensate. The dotted line is the equilibrium $n^{\rm eq}_k[n_{0,f}]$ and the dashed line corresponds to the quasi-adiabatic momentum distribution $n^{\rm eq}_k[n_0(t)]$ (see Eqs. \eqref{eq_neq} and \eqref{eq_n0}). In both figures $\Gamma=\gamma_0 t_0=0.1$.
}
\label{fig_time_ev}
\end{figure*}

\textit{Dynamics of the momentum distribution- }
To discuss the behavior of the momentum density after an interaction
quench, it is convenient to introduce a characteristic time-scale $t_0=(g_f n_{0,f})^{-1}$ corresponding to the characteristic energy scale of a Bose condensate $g_f n_{0,f}$.
We first discuss the case with constant
condensate density $n_0(t)=n_{0,i}=n_{0,f}$, which is appropriate for small quenches
\cite{Natu2013,Hung2013,Rancon2013b}.  Fig.
\ref{fig_energy} shows the momentum density as a function of
$\epsilon_k$ at different times. One observes that at smaller
momentum, $k$-dependent
oscillations appear after sufficient wait-times, but at larger
$k$, there is no perceptible time dependence; the system has
equilibrated.
(We cannot extract the Tan
contact 
\cite{Tan2008}
from the large $k$ tails, since the bath model treats
high energy states as a dissipation mechanism.)

The inset
shows the time evolution of the momentum distribution for a typical
$\Gamma=0.1$ \cite{Rancon2013b} as well as the results obtained from
Bogoliubov theory \cite{Natu2013}. One sees that without
dissipation, the momentum distribution, including the Tan contact, has unphysical undamped
oscillations, as 
reported by other groups
(who also introduced a time dependent condensate)
\cite{Yin2013,Sykes2014}; it was argued that,
because these oscillations disappear upon integrating over momenta,
they are less problematic; here we maintain that these integrated quantities are not 
representative
of a metastable state.

In order to illustrate the effects of a time-varying condensate,
we choose, $n_{0,i}=2 n_{0,f}$ to correspond to
roughly the depletion which can be extrapolated from Ref. \onlinecite{Makotyn2014}. The time-scales  of the condensate $\gamma_0^{-1}$ and
of the excitation 
$\tau_k = \frac{1} {\Gamma ( \epsilon_k + g n_0)}$ are chosen such that
(for definiteness) $\Gamma = \gamma_0/g_f n_0 = 0.1$.
The left panel of Fig. \ref{fig_time_ev} compares the momentum distributions (solid
curves) for two different
$k$. 
The arrows indicate the characteristic time
$\tau_{\rm interm}(k)$. 

That the solid curves in the left panel of Fig.~\ref{fig_time_ev} are still far from their
long time asymptotes is illustrated
through a
blow up of the lower curve
in the right
panel of Fig.~\ref{fig_time_ev}. One sees that
the momentum density has a very slow, non-oscillatory, dynamics, reflecting
the instantaneous value of the condensate $n_0(t)$;  that is, 
 $n_k(t)\simeq n^{\rm eq}_k[n_0(t)]$, where 
\begin{equation}
n^{\rm eq}_k[n_0]=\frac{1}{2}\left(\frac{\epsilon_k+g_f n_0}{\sqrt{(\epsilon_k(\epsilon_k+2 g_f n_0)}}-1\right)
\label{eq_neq}
\end{equation}
 is the equilibrium value of the momentum distribution for a condensate density $n_0$ (and interaction strength $g_f$). 
This is a quasi-adiabatic process in which the
large $k$ states are able to 
follow the condensate in time. Nevertheless until the condensate
has reached it final value, the system is not fully equilibrated.

We comment now on the relation to the experimental data 
from
Ref. \onlinecite{Makotyn2014}
which was
presented in 
Figure \ref{fig_tk}.
Our results support the observation of these authors that ``the higher momentum population
saturates earlier".

Given that we have argued there are multiple time-scales, it is
important to infer which of these is represented
by their data in the inset.
With the caveat that our Bogoliubov-based theory may not be
relevant to quenches to unitarity, we can nevertheless infer from
their Figure 4, that
experimentally,
on the time-scales studied, the measured $n_{k}$ appears to be
time dependent at small ${k}$. This suggests that the relevant measurement times
correspond to $\tau_{\rm interm}\ll t\ll \tau_{\rm long}$.

We stress that
the $k$ units used in our figures and that of Ref.~\onlinecite{Makotyn2014}  are of the same order of magnitude when the energy $\eps_k$ is normalized using the density as only length scale. To see this note that
 $ g_f n_{0,f}=\epsilon_{k_h}=k_h^2/2m$ corresponds to the typical
 kinetic energy of the condensate after the quench ($k_h$ is the inverse healing
 length). At large s-wave scattering length $a$, the interaction strength $g_f=4\pi a/m$ is not well defined. As in Ref.~\onlinecite{Yin2013}, one replaces in this case $g_f$ by an effective T-matrix $\tilde
 g_f=\frac{4\pi a/m}{1+\alpha a n^{1/3}}$ with $n$ the density and $\alpha$ a
numerical constant. At unitarity, one obtains $m \tilde g_f\propto n^{-1/3}$, and
therefore $\tilde g_f n\propto n^{2/3}/m$ is of the order of $\epsilon_n=\frac{(
6\pi2 n)^{2/3}}{2m}$ (assuming $n_0\simeq n$). Thus the energy (momentum) range
 of Fig. \ref{fig_tk} and that of the figure  of Ref. \onlinecite
{Makotyn2014} shown in the inset are of the same order of magnitude.


\textit{Conclusion-}
In summary, in this paper, 
we have addressed the various time scales and the nature of the equilibration
process in three dimensional Bose gases. 
We stress that these
calculations, based on
a bath approach, 
lead to characteristic time-scales which are quite general.
They can be
extracted for instance from the dissipative Gross-Pitaevskii scheme at 
a heuristic level.

Also intuitive should be our major conclusion which follows from the 
experimental observations
\cite{Makotyn2014,Hung2013}
that large $k$ excited states equilibrate most rapidly. This
leads to
an interesting phenomenon in which the system may \textit{appear} to be equilibrated
(at large $k$),
even though it is not. 
This quasi-equilibrated
phase corresponds to a situation in which the high energy (large momentum) excitations, 
which are damped out more rapidly than the condensate,
are able to adiabatically follow the time evolution of the condensate.
It is only when the condensate has reached its final time-independent
state, at $t\gg \gamma_0^{-1}$, that   
full equilibration is reached.

Recent experiments have provided the first glimpse of
unitary
Bose gases
\cite{Makotyn2014} formed through a quench. Figure 4 from their
paper suggests that the small momentum states may not have equilibrated.
While the
present approach is restricted to a Bogoliubov-based scheme, 
our work suggests that
the time scales of this important experiment may
correspond to
$\tau_{\rm interm}(k)$.

\vskip2mm

We thank K. Hazzard, J. Corson and C.-L. Hung for useful discussions, as well as P. Makotyn for sharing 
the experimental data of Ref. \onlinecite{Makotyn2014}.
This work is supported by NSF-MRSEC Grant
0820054.


%

\section*{Supplemental materials}

We give here the technical details of our bath approach to the dynamics of a quench Bose gas that was developed in Ref.~\onlinecite{Rancon2013b}.
Following Ref.~\onlinecite{Rancon2013b}, we compute the momentum distribution $n_k(t)$.
Based on Bogoliubov theory, we consider
the full Hamiltonian in the absence of a trap to be given by
$\hat H(g;n_0)=\hat H_{{\rm Bog}}(g;n_0) +
\hat H_{{\rm bath}} +
\hat H_{\rm {coup}}$
where
\begin{eqnarray}
 \hat H_{{\rm Bog}}(g;n_0) &=& \sum_k \big[\hat \psi^\dag_k (\eps_k+gn_0) \hat \psi_k + \frac{gn_0}{2} \hat \psi_k\hat \psi_{-k} \nonumber \\
&+& \frac{gn_0}{2} \hat \psi^\dag_k\hat \psi^\dag_{-k}\big] \nonumber \\
\hat H_{{\rm bath}}&=& \sum_{i,k} \big[\omega_{i,k}\hat W^\dag_{i,k}\hat W_{i,k}+\nu_{i,k}\hat V^\dag_{i,k}\hat V_{i,k}\big] \nonumber \\
\hat H_{\rm {coup}}&=& \sum_{i,k} \big[\eta_{i,k}^* \hat W^\dag_{i,k} \hat \psi_k +\zeta_{i,k
} \hat V^\dag_{i,-k} \hat \psi^\dag_k+ h.c.\big] \nonumber 
\end{eqnarray}
where $\hat \psi^{(\dag)}_k$ annihilates (creates) an atom with momentum $k\neq0$ 
Here $n_0$ is the condensate density and $g$ is the interaction strength.
The bath is
characterized by two kinds of bosonic modes, $\hat W^{(\dag)}_{i,k}$ and $\hat V^{(\dag)}_{i,k}$, which allow for a well-behaved spectral function \cite{Tan2004,Rancon2013b}.

The dynamics of the system after an interaction quench from $g_i$ to $g_f$ is described by $i\partial_t \hat \psi_k(t)=\big[\hat \psi_k(t),\hat H\big(g_f,n_0(t)\big)\big]$, etc., where we have allowed a time dependent condensate. One can solve the equation for the bath operators which in turn gives
\begin{equation}
\begin{split}
  i\partial_t\hat \psi_k(t)  =& \w_k(t) \hat \psi_k(t)+ g_f n_0(t) \hat  
\psi^\dag_{-k}(t)+\hat D_k(t)\\ 
&-i \int_{t_0}^{t}ds\,\gamma_k(t-s)\hat\psi_k(s) ,\\
 i\partial_t\hat \psi^\dag_{-k}(t)  =& -\w_k(t) \hat \psi^\dag_{-k}(t)- g_f n_0(t) \hat  
\psi_{k}(t)-\hat D^\dag_{-k}(t)\\ 
&-i \int_{t_0}^{t}ds\,\gamma_k(t-s)\hat\psi^\dag_{-k}(s) .
\label{eq_motion}
\end{split}
\end{equation}
Here,
 $\w_k(t) =\eps_k+g_fn_0(t)$, $\hat D_k(t)=\sum_j \eta_{j,k} e^{-i\w_{j,k}t}\hat W_{j,k}(0)+\sum_j \zeta_{j,k} e^{i\nu_{j,k}t}\hat V^\dag_{j,k}(0)$
 and $\gamma_k(t) =\int_\w \Sigma_2(k,\w) e^{-i\w t}$ with
$\int_\w=\int d\w/(2\pi)$. We define the spectral function of the bath
$\Sigma_2(k,\w)=2\pi\sum_j \Big[|\eta_{j,k}|^2 \delta(\w-\w_{j,k})- |\zeta_{j,k}|^2
 \delta(\w+\nu_{j,k})\Big]$. In the following, we will use an Ohmic bath $\Sigma_2(k,\w)=2\Gamma_k \w f(\w/\Omega)$ where $f(\w/\Omega)$ is an even function that 
regularizes the high-energy behavior with cut-off $\Omega$ \cite{WeissBook}. Note that in this framework, the bath parameter $\Gamma_k$ is independent of the temperature of the bath (it is only related to the microscopic coupling between the bath and the system).
 
Note that $\hat D_k(t)$ plays the role of a random force operator and $\gamma_k(t)$ 
reflects the damping. The relaxation to equilibrium will be insured by the satisfaction of
the fluctuation-dissipation relation 
\begin{equation}
 \Big[\hat D_k(t),\hat D^\dag_k(s)\Big]=\gamma_k(t-s).
 \end{equation}

The equations of motion 
\eqref{eq_motion} 
can be formally solved by introducing a matrix Green's function 
\begin{equation}
M_k(t,s)= \begin{pmatrix}
        M_{1,k}(t,s) && M_{2,k}(t,s)\\
         M_{3,k}(t,s) && M_{4,k}(t,s)
       \end{pmatrix},
\end{equation}
 where 
  \begin{equation}
  \begin{split}
i\partial_t M_k(t,s)
=& \begin{pmatrix}
\w_k(t)-i\gamma* && g_f n_0(t)\\
-g_f n_0(t) && \w_k(t)-i\gamma*
\end{pmatrix}
M_k(t,s),
\end{split}
\end{equation}
and $\gamma_k*f(t,s)= \int_0^t du\,\gamma_k(t-u) \, f(u,s)$ for any function $f(t,s)$. The initial condition is given by $M_k(t,t)=-i\large\mathds{1}$. One readily shows that $M_{4,k}^*(t,s)=M_{1,k}(t,s)$ $M_{3,k}^*(t,s)=M_{2,k}(t,s)$. The formal solution of Eq. \eqref{eq_motion} can be written as
  \begin{eqnarray}
 \begin{pmatrix}
        \hat\psi_k(t)\\\hat\psi^\dag_{-k}(t)
       \end{pmatrix}
&=& M_k(t,0)\begin{pmatrix}
        i\hat\psi_{k,0}\\ i\hat\psi^\dag_{-k,0}
       \end{pmatrix} \nonumber \\ 
&+&\int_0^{t}ds M_k (t,s)\begin{pmatrix}
        \hat D_k(s)\\ -\hat D^\dag_{-k}(s)
       \end{pmatrix}.  \label{eq_ev}
\end{eqnarray}
Eq. \eqref{eq_ev} is a generalization of the time-dependent Bogoliubov-de Gennes equation, that includes both dissipation and equilibration.
For a time-dependent condensate, 
$M_k(t,s)$ depends on two times separately, whereas it is a function of $t-s$ if the condensate is constant, as was studied in Ref. \onlinecite{Rancon2013b}.
Here one has to solve the time
evolution matrix numerically when the time dependence of the condensate is specified (see main text).

To compute an observable such as the momentum distribution $n_k=\mean{ \hat\psi^\dag_{k}(t)\hat\psi_k(t)}$, one has to specify the initial state of the system, through the initial correlation functions $\mean{i\hat\psi^{(\dag)}_{\pm k,0}\hat\psi^{(\dag)}_{\pm k,0}}$, $\mean{ \hat\psi^{(\dag)}_{\pm k,0}\hat W^{(\dag)}_{j,\pm k}(0)}$, \emph{etc}. In order to simplify both the discussion and the numerical calculations, we will assume that at $t=0$, the system is an ideal Bose gas ($n_0(t=0^-)=n_{0,i}$ and $g_i=0$) that does not interact with the bath, 
leading to the simplification that all cross-correlation functions such as $\mean{\hat \psi_{k,0} \hat W_{j,k}(0)}$ vanish. We will furthermore assume that $\Gamma_k=\Gamma$ is momentum independent.
For simplicity, the bath is assumed to be at zero temperature. Then, the momentum distribution is given by 
 \begin{equation}
\begin{split}
 n_k(t)=&-M_{k,3}(t,0)M_{k,2}(t,0)
 			-\int_0^t ds\,\int_0^t du\,\calD_k(s-u) \\& \Big[M_{k,1}(t,u)M_{k,4}(t,s)+M_{k,2}(t,u)M_{k,3}(t,s)\Big],
 \end{split}
 \label{eq_nk}
\end{equation}
where $\calD_k(s-u)=\mean{\hat D^\dag_k(s)\hat D_k(u)}=\mean{\hat D_k(s)\hat D^\dag_k(u)}$ is given by $\calD_k(s-u)=\int_{\w<0} \Sigma_2(k,\w) e^{i\w(s-u)}$. In the limit of vanishing system-bath coupling, we 
obtain the standard Bogoliubov results \cite{Natu2013}.

\bibliography{extra_bib,/home/rancon/Dropbox/Articles/bibli_bosons,/home/rancon/Dropbox/Articles/bibli_RG,/home/rancon/Dropbox/Articles/bibli_disorder,/home/rancon/Dropbox/Articles/bibli_OutEq,/home/rancon/Dropbox/Articles/bibli_BCSBEC,/home/rancon/Dropbox/Articles/bibli_diverse}

\begin{thebibliography}{10}%
\makeatletter
\providecommand \@ifxundefined [1]{%
 \ifx #1\undefined \expandafter \@firstoftwo
 \else \expandafter \@secondoftwo
\fi
}%
\providecommand \@ifnum [1]{%
 \ifnum #1\expandafter \@firstoftwo
 \else \expandafter \@secondoftwo
\fi
}%
\providecommand \enquote [1]{``#1''}%
\providecommand \bibnamefont  [1]{#1}%
\providecommand \bibfnamefont [1]{#1}%
\providecommand \citenamefont [1]{#1}%
\providecommand\href[0]{\@sanitize\@href}%
\providecommand\@href[1]{\endgroup\@@startlink{#1}\endgroup\@@href}%
\providecommand\@@href[1]{#1\@@endlink}%
\providecommand \@sanitize [0]{\begingroup\catcode`\&12\catcode`\#12\relax}%
\@ifxundefined \pdfoutput {\@firstoftwo}{%
 \@ifnum{\z@=\pdfoutput}{\@firstoftwo}{\@secondoftwo}%
}{%
 \providecommand\@@startlink[1]{\leavevmode}%
 \providecommand\@@endlink[0]{}%
}{%
 \providecommand\@@startlink[1]{%
  \leavevmode
  \pdfstartlink
   attr{/Border[0 0 1 ]/H/I/C[0 1 1]}%
   user{/Subtype/Link/A<</Type/Action/S/URI/URI(#1)>>}%
  \relax
 }%
 \providecommand\@@endlink[0]{\pdfendlink}%
}%
\providecommand \url  [0]{\begingroup\@sanitize \@url }%
\providecommand \@url [1]{\endgroup\@href {#1}{\urlprefix}}%
\providecommand \urlprefix [0]{URL }%
\providecommand \Eprint[0]{\href }%
\@ifxundefined \urlstyle {%
  \providecommand \doi [1]{doi:\discretionary{}{}{}#1}%
}{%
  \providecommand \doi [0]{doi:\discretionary{}{}{}\begingroup
  \urlstyle{rm}\Url }%
}%
\providecommand \doibase [0]{http://dx.doi.org/}%
\providecommand \Doi[1]{\href{\doibase#1}}%
\providecommand \bibAnnote [3]{%
  \BibitemShut{#1}%
  \begin{quotation}\noindent
    \textsc{Key:}\ #2\\\textsc{Annotation:}\ #3%
  \end{quotation}%
}%
\providecommand \bibAnnoteFile [2]{%
  \IfFileExists{#2}{\bibAnnote {#1} {#2} {\input{#2}}}{}%
}%
\providecommand \typeout [0]{\immediate \write \m@ne }%
\providecommand \selectlanguage [0]{\@gobble}%
\providecommand \bibinfo [0]{\@secondoftwo}%
\providecommand \bibfield [0]{\@secondoftwo}%
\providecommand \translation [1]{[#1]}%
\providecommand \BibitemOpen[0]{}%
\providecommand \bibitemStop [0]{}%
\providecommand \bibitemNoStop [0]{.\EOS\space}%
\providecommand \EOS [0]{\spacefactor3000\relax}%
\providecommand \BibitemShut [1]{\csname bibitem#1\endcsname}%
\bibitem{Donley2001}%
  \BibitemOpen
  \bibfield{author}{%
  \bibinfo {author} {\bibfnamefont{E.~A.}\ \bibnamefont{Donley}}, \bibinfo
  {author} {\bibfnamefont{N.~R.}\ \bibnamefont{Claussen}}, \bibinfo {author}
  {\bibfnamefont{S.~L.}\ \bibnamefont{Cornish}}, \bibinfo {author}
  {\bibfnamefont{J.~L.}\ \bibnamefont{Roberts}}, \bibinfo {author}
  {\bibfnamefont{E.~A.}\ \bibnamefont{Cornell}},\ and\ \bibinfo {author}
  {\bibfnamefont{C.~E.}\ \bibnamefont{Wieman}},\ }%
  \bibfield{journal}{%
  \bibinfo {journal} {Nature}\ }%
  \textbf{\bibinfo {volume} {412}},\ \bibinfo {pages} {295} (\bibinfo {year} {2001})%
  \bibAnnoteFile{NoStop}{Donley2001}%
\bibitem{Greiner2005}%
  \BibitemOpen
  \bibfield{author}{%
  \bibinfo {author} {\bibfnamefont{M.}~\bibnamefont{Greiner}}, \bibinfo
  {author} {\bibfnamefont{C.~A.}\ \bibnamefont{Regal}},\ and\ \bibinfo {author}
  {\bibfnamefont{D.~S.}\ \bibnamefont{Jin}},\ }%
  \bibfield{journal}{%
  \Doi{10.1103/PhysRevLett.94.070403}{\bibinfo {journal} {Phys. Rev. Lett.}}\
  }%
  \textbf{\bibinfo {volume} {94}},\ \bibinfo {pages} {070403} (\bibinfo {year} {2005})%
  \bibAnnoteFile{NoStop}{Greiner2005}%
\bibitem{Smith2012}%
  \BibitemOpen
  \bibfield{author}{%
  \bibinfo {author} {\bibfnamefont{R.~P.}\ \bibnamefont{Smith}}, \bibinfo
  {author} {\bibfnamefont{S.}~\bibnamefont{Beattie}}, \bibinfo {author}
  {\bibfnamefont{S.}~\bibnamefont{Moulder}}, \bibinfo {author}
  {\bibfnamefont{R.~L.~D.}\ \bibnamefont{Campbell}},\ and\ \bibinfo {author}
  {\bibfnamefont{Z.}~\bibnamefont{Hadzibabic}},\ }%
  \bibfield{journal}{%
  \Doi{10.1103/PhysRevLett.109.105301}{\bibinfo {journal} {Phys. Rev. Lett.}}\
  }%
  \textbf{\bibinfo {volume} {109}},\ \bibinfo {pages} {105301} (\bibinfo {year} {2012})%
  \bibAnnoteFile{NoStop}{Smith2012}%
\bibitem{Hung2013}%
  \BibitemOpen
  \bibfield{author}{%
  \bibinfo {author} {\bibfnamefont{C.-L.}\ \bibnamefont{Hung}}, \bibinfo
  {author} {\bibfnamefont{V.}~\bibnamefont{Gurarie}},\ and\ \bibinfo {author}
  {\bibfnamefont{C.}~\bibnamefont{Chin}},\ }%
  \bibfield{journal}{%
  \bibinfo {journal} {Science}}%
  \textbf{\bibinfo {volume} { 341}},\ \bibinfo {pages} {1213} (\bibinfo {year} {2013})%
  \bibAnnoteFile{NoStop}{Hung2013}%
\bibitem{Makotyn2014}%
  \BibitemOpen
  \bibfield{author}{%
  \bibinfo {author} {\bibfnamefont{P.}~\bibnamefont{Makotyn}}, \bibinfo
  {author} {\bibfnamefont{C.~E.}\ \bibnamefont{Klauss}}, \bibinfo {author}
  {\bibfnamefont{D.~L.}\ \bibnamefont{Goldberger}}, \bibinfo {author}
  {\bibfnamefont{E.~A.}\ \bibnamefont{Cornell}},\ and\ \bibinfo {author}
  {\bibfnamefont{D.~S.}\ \bibnamefont{Jin}},\ }%
  \bibfield{journal}{%
  \bibinfo {journal} {Nat Phys}\ }%
  \textbf{\bibinfo {volume} {10}},\ \bibinfo {pages} {116} (\bibinfo {year} {2014})%
  \bibAnnoteFile{NoStop}{Makotyn2014}%
\bibitem{Barankov2004}%
  \BibitemOpen
  \bibfield{author}{%
  \bibinfo {author} {\bibfnamefont{R.~A.}\ \bibnamefont{Barankov}}, \bibinfo
  {author} {\bibfnamefont{L.~S.}\ \bibnamefont{Levitov}},\ and\ \bibinfo
  {author} {\bibfnamefont{B.~Z.}\ \bibnamefont{Spivak}},\ }%
  \bibfield{journal}{%
  \Doi{10.1103/PhysRevLett.93.160401}{\bibinfo {journal} {Phys. Rev. Lett.}}\
  }%
  \textbf{\bibinfo {volume} {93}},\ \bibinfo {pages} {160401} (\bibinfo {year} {2004})%
  \bibAnnoteFile{NoStop}{Barankov2004}%
\bibitem{Altschuler2}%
  \BibitemOpen
  \bibfield{author}{%
  \bibinfo {author} {\bibfnamefont{E.~A.}\ \bibnamefont{Yuzbashyan}}, \bibinfo
  {author} {\bibfnamefont{O.}~\bibnamefont{Tsyplyatyev}},\ and\ \bibinfo
  {author} {\bibfnamefont{B.~L.}\ \bibnamefont{Altshuler}},\ }%
  \bibfield{journal}{%
  \bibinfo {journal} {Phys. Rev. Lett.}\ }%
  \textbf{\bibinfo {volume} {96}},\ \bibinfo {pages} {097005} (\bibinfo {year}
  {2006})%
  \bibAnnoteFile{NoStop}{Altschuler2}%
\bibitem{Natu2013}%
  \BibitemOpen
  \bibfield{author}{%
  \bibinfo {author} {\bibfnamefont{S.~S.}\ \bibnamefont{Natu}}\ and\ \bibinfo
  {author} {\bibfnamefont{E.~J.}\ \bibnamefont{Mueller}},\ }%
  \bibfield{journal}{%
  \Doi{10.1103/PhysRevA.87.053607}{\bibinfo {journal} {Phys. Rev. A}}\ }%
  \textbf{\bibinfo {volume} {87}},\ \bibinfo {pages} {053607} (\bibinfo {year} {2013})%
  \bibAnnoteFile{NoStop}{Natu2013}%
\bibitem{Yin2013}%
  \BibitemOpen
  \bibfield{author}{%
  \bibinfo {author} {\bibfnamefont{X.}~\bibnamefont{Yin}}\ and\ \bibinfo
  {author} {\bibfnamefont{L.}~\bibnamefont{Radzihovsky}},\ }%
  \bibfield{journal}{%
  \Doi{10.1103/PhysRevA.88.063611}{\bibinfo {journal} {Phys. Rev. A}}\ }%
  \textbf{\bibinfo {volume} {88}},\ \bibinfo {pages} {063611} (\bibinfo {year} {2013})%
  \bibAnnoteFile{NoStop}{Yin2013}%
\bibitem{Sykes2014}%
  \BibitemOpen
  \bibfield{author}{%
  \bibinfo {author} {\bibfnamefont{A.~G.}\ \bibnamefont{Sykes}}, \bibinfo
  {author} {\bibfnamefont{J.~P.}\ \bibnamefont{Corson}}, \bibinfo {author}
  {\bibfnamefont{J.~P.}\ \bibnamefont{D'Incao}}, \bibinfo {author}
  {\bibfnamefont{A.~P.}\ \bibnamefont{Koller}}, \bibinfo {author}
  {\bibfnamefont{C.~H.}\ \bibnamefont{Greene}}, \bibinfo {author}
  {\bibfnamefont{A.~M.}\ \bibnamefont{Rey}}, \bibinfo {author}
  {\bibfnamefont{K.~R.~A.}\ \bibnamefont{Hazzard}},\ and\ \bibinfo {author}
  {\bibfnamefont{J.~L.}\ \bibnamefont{Bohn}},\ }%
  \bibfield{journal}{%
  \Doi{10.1103/PhysRevA.89.021601}{\bibinfo {journal} {Phys. Rev. A}}\ }%
  \textbf{\bibinfo {volume} {89}},\ \bibinfo {pages} {021601} (\bibinfo {year} {2014})%
  \bibAnnoteFile{NoStop}{Sykes2014}%
\bibitem{Kain2014}%
  \BibitemOpen
  \bibfield{author}{%
  \bibinfo {author} {\bibfnamefont{B.}~\bibnamefont{Kain}}\ and\ \bibinfo
  {author} {\bibfnamefont{H.~Y.}\ \bibnamefont{Ling}},\ }%
  \bibfield{journal}{%
  \bibinfo {journal} {arXiv:1401.2390 }}%
   (\bibinfo {year} {2014})%
  \bibAnnoteFile{NoStop}{Kain2014}%
\bibitem{arxiv}%
  \BibitemOpen
  \bibinfo {journal} {arXiv: 1209.0011}%
  \bibAnnoteFile{NoStop}{arxiv}%
\bibitem{Rancon2013b}%
  \BibitemOpen
\bibfield{journal}{%
    }%
  \bibfield{author}{%
  \bibinfo {author} {\bibfnamefont{A.}~\bibnamefont{Ran\c{c}on}}, \bibinfo
  {author} {\bibfnamefont{C.-L.}\ \bibnamefont{Hung}}, \bibinfo {author}
  {\bibfnamefont{C.}~\bibnamefont{Chin}},\ and\ \bibinfo {author}
  {\bibfnamefont{K.}~\bibnamefont{Levin}},\ }%
  \bibfield{journal}{%
  \Doi{10.1103/PhysRevA.88.031601}{\bibinfo {journal} {Phys. Rev. A}}\ }%
  \textbf{\bibinfo {volume} {88}},\ \bibinfo {pages} {031601} (\bibinfo {year} {2013})%
  \bibAnnoteFile{NoStop}{Rancon2013b}%
\bibitem{Choi1998}%
  \BibitemOpen
  \bibfield{author}{%
  \bibinfo {author} {\bibfnamefont{S.}~\bibnamefont{Choi}}, \bibinfo {author}
  {\bibfnamefont{S.~A.}\ \bibnamefont{Morgan}},\ and\ \bibinfo {author}
  {\bibfnamefont{K.}~\bibnamefont{Burnett}},\ }%
  \bibfield{journal}{%
  \Doi{10.1103/PhysRevA.57.4057}{\bibinfo {journal} {Phys. Rev. A}}\ }%
  \textbf{\bibinfo {volume} {57}},\ \bibinfo {pages} {4057} (\bibinfo {year} {1998})%
  \bibAnnoteFile{NoStop}{Choi1998}%
\bibitem{Stoof1999}%
  \BibitemOpen
  \bibfield{author}{%
  \bibinfo {author} {\bibfnamefont{H.~T.~C.}\ \bibnamefont{Stoof}},\ }%
  \bibfield{journal}{%
  \bibinfo {journal} {Journal of Low Temperature Physics}\ }%
  \textbf{\bibinfo {volume} {114}},\ \bibinfo {pages} {11} (\bibinfo {year}
  {1999})%
  \bibAnnoteFile{NoStop}{Stoof1999}%
\bibitem{Zaremba1999}%
  \BibitemOpen
  \bibfield{author}{%
  \bibinfo {author} {\bibfnamefont{E.}~\bibnamefont{Zaremba}}, \bibinfo
  {author} {\bibfnamefont{T.}~\bibnamefont{Nikuni}},\ and\ \bibinfo {author}
  {\bibfnamefont{A.}~\bibnamefont{Griffin}},\ }%
  \bibfield{journal}{%
  \Doi{10.1023/A:1021846002995}{\bibinfo {journal} {Journal of Low Temperature
  Physics}}\ }%
  \textbf{\bibinfo {volume} {116}},\ \bibinfo {pages} {277} (\bibinfo {year}
  {1999})%
  \bibAnnoteFile{NoStop}{Zaremba1999}%
\bibitem{Blakie2008}%
  \BibitemOpen
  \bibfield{author}{%
  \bibinfo {author} {\bibfnamefont{P.}~\bibnamefont{Blakie}}, \bibinfo {author}
  {\bibfnamefont{A.}~\bibnamefont{Bradley}}, \bibinfo {author}
  {\bibfnamefont{M.}~\bibnamefont{Davis}}, \bibinfo {author}
  {\bibfnamefont{R.}~\bibnamefont{Ballagh}},\ and\ \bibinfo {author}
  {\bibfnamefont{C.}~\bibnamefont{Gardiner}},\ }%
  \bibfield{journal}{%
  \Doi{10.1080/00018730802564254}{\bibinfo {journal} {Advances in Physics}}\ }%
  \textbf{\bibinfo {volume} {57}},\ \bibinfo {pages} {363} (\bibinfo {year}
  {2008})%
  \bibAnnoteFile{NoStop}{Blakie2008}%
\bibitem{Note1}%
  \BibitemOpen
  \bibinfo {note} {In all rigor, we find $\tau _{\protect \rm short}(k) \propto
  \protect \frac {1}{\protect \sqrt {E_k2-(\gamma gn_0)2 }}$. For momentum
  smaller than a cross-over momentum $ k_x$ (defined such that $E_{k_x} \equiv
  \gamma gn_0$), the dynamics of the excitations is very different from that of
  (undamped) Bogoliubov theory.}%
  \bibAnnoteFile{Stop}{Note1}%
\bibitem{Tan2008}%
  \BibitemOpen
  \bibfield{author}{%
  \bibinfo {author} {\bibfnamefont{S.}~\bibnamefont{Tan}},\ }%
  \bibfield{journal}{%
  \Doi{http://dx.doi.org/10.1016/j.aop.2008.03.005}{\bibinfo {journal} {Annals
  of Physics}}\ }%
  \textbf{\bibinfo {volume} {323}},\ \bibinfo {pages} {2971 } (\bibinfo {year}
  {2008})%
  \bibAnnoteFile{NoStop}{Tan2008}%
\bibitem{Caldeira1983}%
  \BibitemOpen
  \bibfield{author}{%
  \bibinfo {author} {\bibfnamefont{A.}~\bibnamefont{Caldeira}}\ and\ \bibinfo
  {author} {\bibfnamefont{A.}~\bibnamefont{Leggett}},\ }%
  \bibfield{journal}{%
  \Doi{10.1016/0378-4371(83)90013-4}{\bibinfo {journal} {Physica A}}\ }%
  \textbf{\bibinfo {volume} {121}},\ \bibinfo {pages} {587 } (\bibinfo {year}
  {1983})%
  \bibAnnoteFile{NoStop}{Caldeira1983}%
\bibitem{WeissBook}%
  \BibitemOpen
  \bibfield{author}{%
  \bibinfo {author} {\bibfnamefont{U.}~\bibnamefont{Weiss}},\ }%
  \emph{\bibinfo {title} {Quantum Dissipative Systems}}\ (\bibinfo {publisher}
  {World Scientific},\ \bibinfo {year} {1999})%
  \bibAnnoteFile{NoStop}{WeissBook}%
\bibitem{Note2}%
  \BibitemOpen
  \bibinfo {note} {More generally, these and the measurements of $\tau (k)$ in
  Ref. \protect \rev@citealp {Makotyn2014} seem compatible with $\Gamma $
  between $0.1$ and $1.0$.}%
  \bibAnnoteFile{Stop}{Note2}%
\end{thebibliography}

\begin{thebibliography}{1}%
\makeatletter
\providecommand \@ifxundefined [1]{%
 \ifx #1\undefined \expandafter \@firstoftwo
 \else \expandafter \@secondoftwo
\fi
}%
\providecommand \@ifnum [1]{%
 \ifnum #1\expandafter \@firstoftwo
 \else \expandafter \@secondoftwo
\fi
}%
\providecommand \enquote [1]{``#1''}%
\providecommand \bibnamefont  [1]{#1}%
\providecommand \bibfnamefont [1]{#1}%
\providecommand \citenamefont [1]{#1}%
\providecommand\href[0]{\@sanitize\@href}%
\providecommand\@href[1]{\endgroup\@@startlink{#1}\endgroup\@@href}%
\providecommand\@@href[1]{#1\@@endlink}%
\providecommand \@sanitize [0]{\begingroup\catcode`\&12\catcode`\#12\relax}%
\@ifxundefined \pdfoutput {\@firstoftwo}{%
 \@ifnum{\z@=\pdfoutput}{\@firstoftwo}{\@secondoftwo}%
}{%
 \providecommand\@@startlink[1]{\leavevmode}%
 \providecommand\@@endlink[0]{}%
}{%
 \providecommand\@@startlink[1]{%
  \leavevmode
  \pdfstartlink
   attr{/Border[0 0 1 ]/H/I/C[0 1 1]}%
   user{/Subtype/Link/A<</Type/Action/S/URI/URI(#1)>>}%
  \relax
 }%
 \providecommand\@@endlink[0]{\pdfendlink}%
}%
\providecommand \url  [0]{\begingroup\@sanitize \@url }%
\providecommand \@url [1]{\endgroup\@href {#1}{\urlprefix}}%
\providecommand \urlprefix [0]{URL }%
\providecommand \Eprint[0]{\href }%
\@ifxundefined \urlstyle {%
  \providecommand \doi [1]{doi:\discretionary{}{}{}#1}%
}{%
  \providecommand \doi [0]{doi:\discretionary{}{}{}\begingroup
  \urlstyle{rm}\Url }%
}%
\providecommand \doibase [0]{http://dx.doi.org/}%
\providecommand \Doi[1]{\href{\doibase#1}}%
\providecommand \bibAnnote [3]{%
  \BibitemShut{#1}%
  \begin{quotation}\noindent
    \textsc{Key:}\ #2\\\textsc{Annotation:}\ #3%
  \end{quotation}%
}%
\providecommand \bibAnnoteFile [2]{%
  \IfFileExists{#2}{\bibAnnote {#1} {#2} {\input{#2}}}{}%
}%
\providecommand \typeout [0]{\immediate \write \m@ne }%
\providecommand \selectlanguage [0]{\@gobble}%
\providecommand \bibinfo [0]{\@secondoftwo}%
\providecommand \bibfield [0]{\@secondoftwo}%
\providecommand \translation [1]{[#1]}%
\providecommand \BibitemOpen[0]{}%
\providecommand \bibitemStop [0]{}%
\providecommand \bibitemNoStop [0]{.\EOS\space}%
\providecommand \EOS [0]{\spacefactor3000\relax}%
\providecommand \BibitemShut [1]{\csname bibitem#1\endcsname}%
\bibitem{Rancon2013b}%
  \BibitemOpen
  \bibfield{author}{%
  \bibinfo {author} {\bibfnamefont{A.}~\bibnamefont{Ran\c{c}on}}, \bibinfo
  {author} {\bibfnamefont{C.-L.}\ \bibnamefont{Hung}}, \bibinfo {author}
  {\bibfnamefont{C.}~\bibnamefont{Chin}},\ and\ \bibinfo {author}
  {\bibfnamefont{K.}~\bibnamefont{Levin}},\ }%
  \bibfield{journal}{%
  \Doi{10.1103/PhysRevA.88.031601}{\bibinfo {journal} {Phys. Rev. A}}\ }%
  \textbf{\bibinfo {volume} {88}},\ \bibinfo {pages} {031601} (\bibinfo {year} {2013})
  \bibAnnoteFile{NoStop}{Rancon2013b}%
\bibitem{Tan2004}%
  \BibitemOpen
  \bibfield{author}{%
  \bibinfo {author} {\bibfnamefont{S.}~\bibnamefont{Tan}}\ and\ \bibinfo
  {author} {\bibfnamefont{K.}~\bibnamefont{Levin}},\ }%
  \bibfield{journal}{%
  \Doi{10.1103/PhysRevB.69.064510}{\bibinfo {journal} {Phys. Rev. B}}\ }%
  \textbf{\bibinfo {volume} {69}},\ \bibinfo {pages} {064510} (\bibinfo {year} {2004})
   \bibAnnoteFile{NoStop}{Tan2004}%
\bibitem{WeissBook}%
  \BibitemOpen
  \bibfield{author}{%
  \bibinfo {author} {\bibfnamefont{U.}~\bibnamefont{Weiss}},\ }%
  \emph{\bibinfo {title} {Quantum Dissipative Systems}}\ (\bibinfo {publisher}
  {World Scientific},\ \bibinfo {year} {1999})%
  \bibAnnoteFile{NoStop}{WeissBook}%
\bibitem{Natu2013}%
  \BibitemOpen
  \bibfield{author}{%
  \bibinfo {author} {\bibfnamefont{S.~S.}\ \bibnamefont{Natu}}\ and\ \bibinfo
  {author} {\bibfnamefont{E.~J.}\ \bibnamefont{Mueller}},\ }%
  \bibfield{journal}{%
  \Doi{10.1103/PhysRevA.87.053607}{\bibinfo {journal} {Phys. Rev. A}}\ }%
  \textbf{\bibinfo {volume} {87}},\ \bibinfo {pages} {053607} (\bibinfo {year} {2013})
   \bibAnnoteFile{NoStop}{Natu2013}%
\end{thebibliography}

%

\end{document}